\documentclass[usenatbib]{mnras}
\usepackage{epsfig,booktabs,caption,fixltx2e,times,graphicx,amsmath}
\usepackage[flushleft]{threeparttable}

\def\Hy@FixNotFirstPage{%
	\gdef\Hy@FixNotFirstPage{%
		\setbox\AtBeginShipoutBox=\hbox{%
			\copy\AtBeginShipoutBox
		}%
	}%
}
\AtBeginShipout{\Hy@FixNotFirstPage}
 
\def\I{\,\textsc{i}}
\def\II{\,\textsc{ii}}
\def\III{\,\textsc{iii}}

\def\hst{{\it HST}}
\def\swift{{\it Swift}}

\title[Progenitor of SN~2016gkg]{On the Progenitor of the Type IIb Supernova 2016gkg}

\author[Kilpatrick et al.]{Charles D. Kilpatrick$^1$\thanks{Email:
    cdkilpat@ucsc.edu}, Ryan J. Foley$^1$, Louis E. Abramson$^2$, Yen-Chen Pan$^1$, 
	\newauthor Cicero-Xinyu Lu$^2$, Peter Williams$^2$, Tommaso Treu$^2$, Matthew R. Siebert$^1$, 
	\newauthor Christopher D. Fassnacht$^3$, Claire E. Max$^1$\\
	$^1$Department of Astronomy and Astrophysics, University of California, Santa Cruz, CA 95064, USA\\
	$^2$Department of Physics and Astronomy, University of California, Los Angeles, CA 90095, USA\\
	$^3$Department of Physics, University of California, One Shields Avenue, Davis, CA 95616, USA}

\begin{document}
\date{Accepted 0000, Received 0000, in original form 0000}
\pagerange{\pageref{firstpage}--\pageref{lastpage}} \pubyear{2016}
\maketitle
\label{firstpage}

\begin{abstract}
\noindent

We present a detection in pre-explosion \textit{Hubble Space Telescope} (\textit{HST}) imaging of a point source consistent with being the progenitor star of the Type IIb supernova (SN~IIb) 2016gkg. Post-explosion imaging from the Keck Adaptive Optics system was used to perform relative astrometry between the Keck and \hst\ imaging. We identify a single point source in the \hst\ images coincident with the SN position to 0.89-$\sigma$. The \hst\ photometry is consistent with the progenitor star being an A0Ia star with $T=9500$~K and $\log (L/L_{\odot}) = 5.15$. We find that the SN~2016gkg progenitor star appears more consistent with binary than single-star evolutionary models. In addition, early-time light curve data from SN~2016gkg revealed a rapid rise in luminosity within $\sim0.4~\text{days}$ of non-detection limits, consistent with models of the cooling phase after shock break-out. We use these data to determine an explosion date of $20.15$ September 2016 and progenitor star radius of $\log (R/R_{\odot}) = 2.41$, which agrees with photometry from the progenitor star. Our findings are also consistent with detections of other SNe~IIb progenitor stars, although more luminous and bluer than most other examples.

\end{abstract}


\section{INTRODUCTION}\label{sec:introduction}

Any normal core-collapse supernova (CCSN) can yield valuable new insight into SN explosion mechanisms when its progenitor star is detected in pre-explosion imaging. The two canonical examples are SNe~1987A and 1993J, which revealed, among other aspects of SNe and their progenitor stars, that binary evolution is central to SNe \citep{aldering+94}, that stars in the mass range 15--$25~M_{\odot}$ explode as SNe \citep{arnett+89,podsialowski+93,maund+04}, and that mass-transfer can describe both the hydrogen envelopes and circumstellar environments of some SNe \citep[e.g.,][]{weiler+07,morris+07}. Beyond these well-studied examples, roughly $20$ progenitor systems have been identified and statistical studies of the connection between progenitor stars and SNe are now possible \citep{smartt+09a,smartt+15}. Of particular interest is the luminosity and colour distribution of these progenitor stars and inferences about their physical properties. Apart from noteable examples such as SN~1987A and SN~2009ip \citep{arnett87,woosley+87,mauerhan+13}, virtually all SN progenitor stars have $B-V > 0.3$~mag (i.e., $T < 7,300~\text{K}$) and most confirmed progenitor stars are red supergiants \citep[as in][]{smartt+15}. This observation is consistent with predictions of star formation and stellar evolution, which suggest that SNe from lower mass and redder stars should be more common. 

It is therefore of enormous scientific value when SN progenitor stars are detected at the extremes of observed colour and luminosity distributions as these stars can both probe unusual SN explosions found in nature and challenge interpretations of stellar evolution and SN physics.  In particular, several SNe~IIb (SNe with a strong hydrogen lines at early times that are relatively weak at later times, implying a thin hydrogen envelope) have been discovered with progenitor star detections, which appear to span the stellar temperature range from red to blue supergiants \citep[see, e.g., SN~2013df and SN~2008ax;][]{vandyk+14,crockett+08}.  Examples such as SN~1993J \citep[][]{aldering+94} challenge single-star evolution models as their progenitor star colours cannot be matched to the end points of most plausible evolutionary tracks. This discovery has led to the interpretation that at least some SNe~IIb come from binary star systems where mass from the progenitor star has been stripped by a companion \citep[][]{nomoto+93,woosley+94,fox+14}.

In this paper, we discuss SN~2016gkg discovered in NGC~613. This SN was discovered by \citet{otero+16} on 20.18 September (all dates presented herein are UT) and reported in a subsequent detection on 20.54 September by \citet{atel9526}.  Within the $9~\text{hours}$ between these detections, the SN appeared to have brightened by $\sim3~\text{mag}$.  \citet{atel9528} reported a spectroscopic confirmation on 21.9 September that SN~2016gkg was a young Type II SN. Subsequent high-resolution spectroscopy on 25.33 September by \citet{atel9562} found broad H$\alpha$ emission with P-Cygni features that matched SN~1987A around peak magnitude.  On 28.56 September, \citet{atel9573} found in low-resolution spectroscopy that SN~2016gkg more closely resembled a SN~IIb.  Here, we present early-time imaging of SN~2016gkg and a subsequent spectral epoch. We discuss detailed astrometry of the SN at early times, which demonstrate that the position of the SN is consistent with a blue source detected in archival \textit{Hubble Space Telescope} (\hst) Wide Field Planetary-Camera 2 (WFPC2) imaging.  We fit the magnitudes derived from this source to stellar spectra and demonstrate that the best match is an A0Ia star. Based on comparison to single and binary stellar evolution tracks, we show that this star most likely evolved in a binary system.  Finally, we analyse the early-time light curve of SN~2016gkg, which rose rapidly in luminosity within a day after discovery, consistent with predictions from the cooling phase of shock break-out.  We show that the stellar radius derived from this light curve is consistent with the radius of the detected progenitor star. Throughout this paper, we assume a Tully-Fisher distance to NGC~613 of $26.4\pm5.3~\text{Mpc}$, with a corresponding distance modulus of $32.11\pm0.44$ \citep{nasonova+11}, and Milky Way extinction of $A_{V} = 0.053$ \citep{schlafly+11}.   

\section{OBSERVATIONS}\label{sec:observations}

\subsection{Archival Data}

\begin{figure*}
	\includegraphics[width=\textwidth]{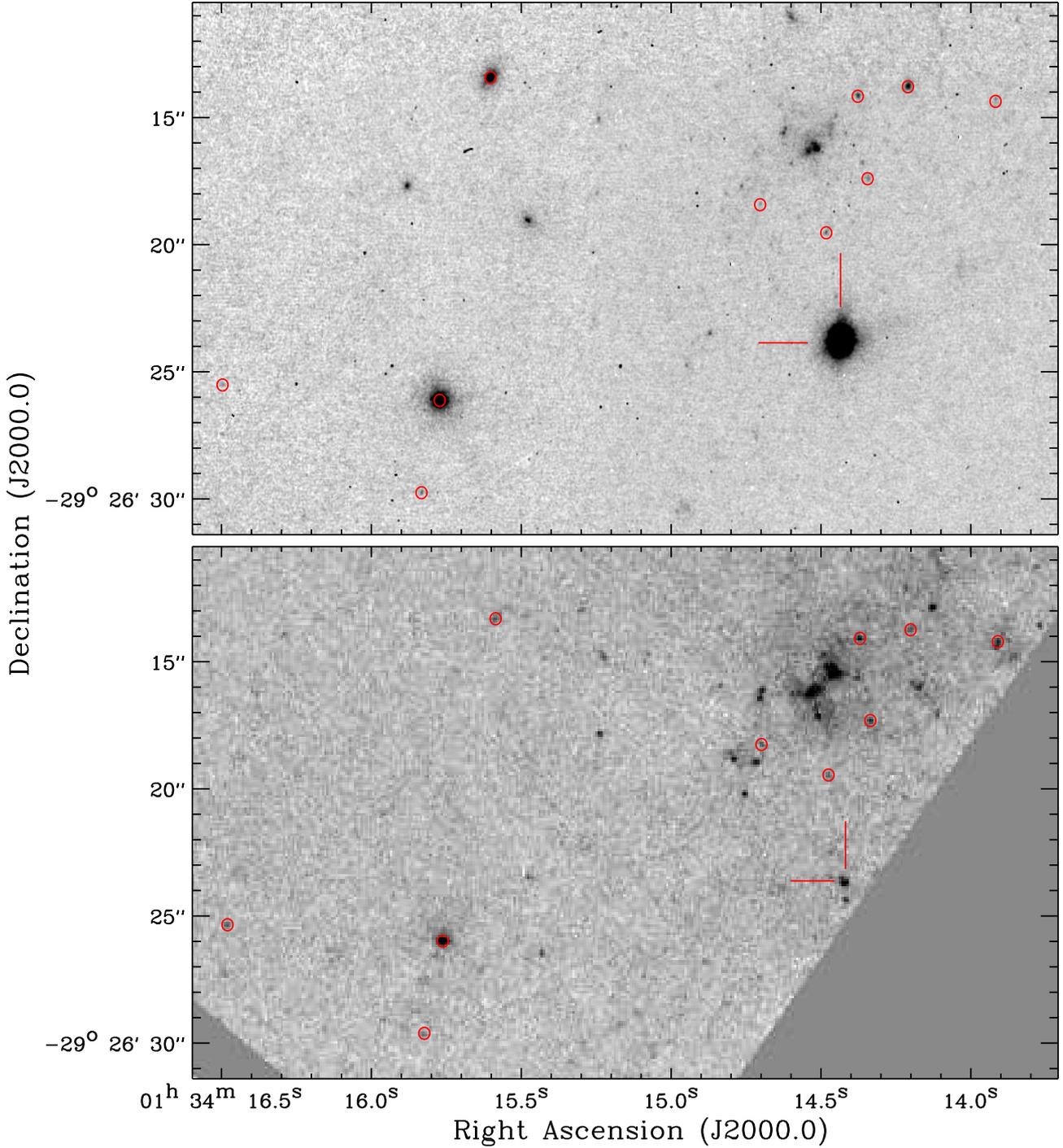}
	\caption{(Upper panel) Keck NIRC2 AO $K^{\prime}$ imaging of SN~2016gkg. The SN is denoted and 10 point sources used for astrometry are circled in red. (Bottom panel) \hst\ WFPC2 F450W+F606W+F814W reference image used for astrometry.  The progenitor star is denoted and the same 10 point sources from the NIRC2 image are circled in red.}\label{fig:astrometry}
\end{figure*}

We obtained archival imaging of NGC~613 from the \hst\ Legacy Archive\footnote{\url{https://hla.stsci.edu/hla_faq.html}} from 21 August 2001 (Cycle 10, Proposal ID 9042, PI Stephen Smartt).  The \hst+WFPC2 data consisted of two frames each of F450W, F606W, and F814W totaling $2\times160~\text{s}$ per filter.  These data had been combined and calibrated by the Canadian Astronomical Data Centre using the latest calibration software and reference files, including corrections for bias, dark current, flat-fielding, and bad pixel masking.  The images had been combined using the \textsc{IRAF}\footnote{IRAF, the Image Reduction and Analysis Facility, is distributed by the National Optical Astronomy Observatory, which is operated by the Association of Universities for Research in Astronomy (AURA) under cooperative agreement with the National Science Foundation (NSF).} task \textsc{MultiDrizzle}, which performs automatic image registration, cosmic ray rejection, and final image combination using the \textsc{Drizzle} task. We performed photometry on these final, calibrated images in each filter using the \textsc{dolphot}\footnote{\url{http://americano.dolphinsim.com/dolphot/}} stellar photometry package.  Finally, we combined the images in all three filters using the \textsc{MultiDrizzle} task, weighting each image by the inverse-variance of emission-free regions in order to produce a reference image with the highest signal-to-noise for each point source.  This image is shown in \autoref{fig:astrometry}.

We also obtained post-explosion photometry of SN~2016gkg recorded from \citet{otero+16,atel9521,atel9526,atel9529}. These data included observations from the All Sky Automated Survey for SuperNovae (ASAS-SN), the Asteroid Terrestrial-impact Last Alert System (ATLAS), and \swift, as well as photometry from the 1-meter telescope on Cerro Tololo, Chile as part of the Las Cumbres Observatory Global Telescope Network (LCOGT).  The early-time photometry is summarized in \autoref{tab:phot}, which was obtained from \citet{atel9529}.

\begin{table}
\begin{center}\begin{minipage}{3.3in}
      \caption{Ultraviolet/Optical Photometry of SN~2016gkg}\scriptsize
\begin{tabular}{@{}cccccc}\hline\hline
  UT Date & Telescope & Filter &Magnitude & Uncertainty & Reference\\ 
  (from 20 September 2016) &           &        &          &             \\   \hline
0.1653   & ASAS-SN   & $V$     &$>$17.36 & ---  & (1)\\
0.2484   & Buso \& Otero &``clear''&17.6 & 0.5      & (2)\\
0.54     &ATLAS   &     $o$&15.94 & 0.13		& (3)\\
0.55     &ATLAS   &     $o$&15.78 & 0.08 & (3)\\
1.1398   & Buso \& Otero &``clear''&14.5 & 0.2  & (2)\\
1.2987   &ASAS-SN  		 &     $V$&15.01 & 0.04 & (1)\\
1.6569   &\swift   		 &    UVW1&13.75 & 0.04 & (4)\\
1.6588   &\swift   		 &     $U$&13.97 & 0.04 & (4)\\
1.6598   &\swift   		 &     $B$&15.21 & 0.04 & (4)\\
1.6608   &\swift   		 &    UVW2&13.92 & 0.04 & (4)\\
1.6645   &\swift   		 &     $V$&15.09 & 0.05 & (4)\\
1.7318   &\swift   		 &    UVM2&15.28 & 0.25 & (4)\\
2.1276   &LCOGT    		 &    $B$ &15.70 & 0.04 & (4)\\
2.1289   &LCOGT    		 &    $V$ &15.54 & 0.03 & (4)\\
2.1302   &LCOGT    		 &    $g$ &15.61 & 0.03 & (4)\\
2.1315   &LCOGT    		 &    $i$ &15.65 & 0.04 & (4)\\
2.1328   &LCOGT    		 &    $r$ &15.59 & 0.04 & (4)\\
2.2884   &ASAS-SN  		 &    $V$ &15.73 & 0.05 & (4)\\
2.3946   &LCOGT    		 &    $B$ &16.01 & 0.08 & (4)\\
2.3959   &LCOGT    		 &    $V$ &15.84 & 0.06 & (4)\\
2.3972   &LCOGT    		 &    $g$ &15.91 & 0.07 & (4)\\
2.3998   &LCOGT    		 &    $r$ &15.85 & 0.08 & (4)\\
\hline
\end{tabular}\label{tab:phot}
\end{minipage}
\end{center}
\begin{tablenotes}
      \small
      \item References: (1) \citet{atel9521}, (2) \citet{otero+16}, (3) \citet{atel9526}, (4) \citet{atel9529}
\end{tablenotes}
\end{table}

\subsection{Adaptive Optics Imaging}

We observed SN~2016gkg in $K^{\prime}$ band with the Near-Infrared Camera 2 (NIRC2) on the Keck-II 10-m telescope in conjunction with the adaptive optics (AO) system on 22 September 2016, as summarized in \citet{atel9536}. These data consisted of 30 individual frames each consisting of 3 co-adds of $10~\text{s}$ for an effective exposure time of $30~\text{s}$ per frame and $900~\text{s}$ total. The individual frames were corrected for pixel-to-pixel variations using a flat-field frame that was created from the science frames themselves, and then sky-subtracted. Images taken with NIRC2 have known optical distortions. Therefore, each of the individual frames was resampled to a corrected grid, using the coordinate distortions that are provided on the NIRC2 website\footnote{\url{http://www2.keck.hawaii.edu/inst/nirc2/nirc2dewarp_positions.pro}}. We masked each individual frame in order to remove bad pixels, cosmic-rays, and additional image artifacts. Finally, we aligned the individual frames using an offset vector calculated from the position of the SN and combined the individual frames. In \autoref{fig:astrometry}, we show the AO imaging along with the reference \hst\ archival image.

\subsection{Spectroscopy}

A spectrum of SN~2016gkg was obtained on 8 October 2016 with the Goodman Spectrograph \citep{clemens+04} and the 4.1-m Southern Astrophysical Research Telescope (SOAR) on Cerro Pach\'{o}n, Chile. We used the 1.07\arcsec\ slit in conjunction with the 400~l/mm grating for an effective spectral range of 4000--7050~\AA\ on the blue side and 5000--9050~\AA\ on the red side and a single $1200~\text{s}$ exposure per side. A blocking filter (GG-455) was used in the red to minimise second-order scattering of blue light onto the CCD. During our observations, we aligned the slit with the center of NGC~613 in order to simultaneously observe the SN and host galaxy. The SN was at an airmass of $\sim$1.01 at this time and chromatic atmospheric dispersion was minimal. Conditions were photometric at the time of observations with $\sim$0.8\arcsec\ seeing. We used \textsc{IRAF} to perform standard reductions on the two-dimensional images and optimal extraction of the one-dimensional blue and red side spectra. We performed wavelength calibration on these one-dimensional images using arc lamp exposures taken immediately after each spectrum.  We derived a sensitivity function from a standard star obtained at similar airmass and in the same instrument configuration and used this function to perform flux calibration. We dereddened the spectrum using the extinction quoted above and removed the recession velocity $v = 1,480~\text{km s}^{-1}$, which is consistent with the velocity of the host galaxy. Finally, we combined the red and blue spectra into a single spectrum, which is presented in \autoref{fig:spectrum}.

\section{RESULTS AND DISCUSSION}\label{sec:results}

\subsection{Spectrum of SN~2016gkg}

\begin{figure}
	\includegraphics[width=0.49\textwidth]{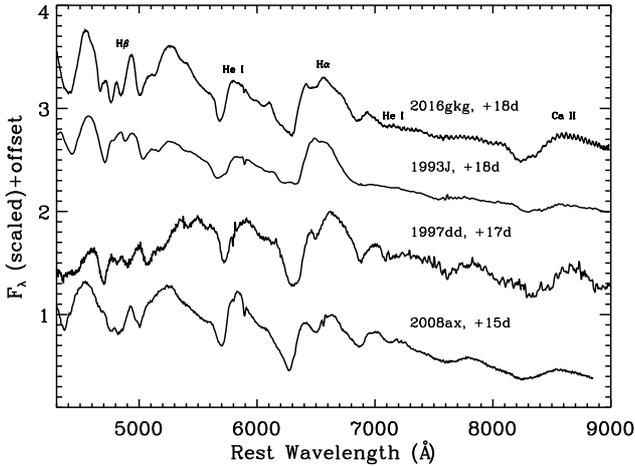}
	\caption{SN~2016gkg spectrum with the day relative to explosion (+\#\#d) of observation given in black. For comparison, we also plot the SNe~IIb 1993J, 1997dd, and 2008ax at a similar epoch relative to explosion \citep[to discovery for 1997dd;][]{matheson+00,matheson+01,taubenberger+11}.  All spectra have been dereddened and their recession velocities have been removed given the parameters provided in each reference. We indicate prominent emission and absorption features in these SN~IIb spectra, including H$\alpha$, H$\beta$, He\I\ $\lambda\lambda$5876 and 7065, and the Ca\II\ infrared triplet.}\label{fig:spectrum}
\end{figure}

In \autoref{fig:spectrum}, we compare our spectrum of SN~2016gkg to spectra of the SNe IIb 1993J, 1997dd, and 2008ax \citep{matheson+00,matheson+01,taubenberger+11}. The comparison spectra have been dereddened and their recession velocities have been removed according to the extinction and redshift information provided in each reference. We indicate the relative epoch of each spectrum with respect to the explosion dates calculated for SNe~2016gkg, 1993J, and 2008ax \citep[see below and][]{matheson+00,taubenberger+11} and with respect to discovery date on 26.18 Aug 1997 for SN~1997dd \citep{nakano+97}.

The comparison between these spectra, especially in He\I\ $\lambda\lambda$5876 and 7065 absorption, strongly suggests that SN~2016gkg is a SN~IIb and was around or slightly before peak magnitude \citep[or secondary peak as in SN~1993J;][]{benson+94}. The development of these He\I\ features appears to be more rapid than in SN~1993J based on our estimated explosion date for SN~2016gkg (\autoref{sec:light-curve}) and is more similar to SN~2008ax at this epoch. In addition, H$\alpha$ is not a dominant emission feature in SN~2016gkg at this epoch, implying that the initial hydrogen emission may be fading relative to the continuum level. Our spectrum is very similar to the early spectrum of SN~1997dd, where a distinct ``notch'' developed in the H$\alpha$ line, likely due to the P-Cygni profile of He\I\ $\lambda$6678 before He absorption had fully developed \citep[e.g.,][]{matheson+01}. We note additional similarities to SN~1997dd, which was identified early as a peculiar Type II SN with weak H$\alpha$ emission \citep{suntzeff+97}, as with SN~2016gkg in analysis by \citet{atel9528}.

As we mention in \autoref{sec:observations}, we aligned the slit of the Goodman Spectrograph to obtain a spectrum of NGC~613 simultaneously with SN~2016gkg.  Analysis of the host galaxy emission line ratio

\begin{equation}
\log R_{23} = \frac{I_{[\text{O} \II]\lambda 3727} + I_{[\text{O}\III]\lambda 4959} + I_{[\text{O}\III]\lambda 5007}}{I_{\text{H} \beta}}
\end{equation}

\noindent using the calibration in \citet{kobulnicky+04} suggests that NGC~613 has $12 + \log (\text{O}/\text{H}) = 8.61\pm0.15$ with an implied metallicity of $Z = 0.012\pm0.004$. This value is slightly sub-solar ($12 + \log (\text{O}/\text{H})_{\odot} = 8.7$), although it agrees with the solar value to within our error bars.  We adopt $Z = 0.012$ for subsequent analysis of the SN~2016gkg progenitor star.

\subsection{Astrometry of the AO Imaging and \hst\ Point Source}

We performed relative astrometry on the AO image and composite \hst\ image using the 10 common sources circled in both frames (\autoref{fig:astrometry}). The positions derived for these 10 sources were determined using \textsc{dolphot} in each frame and image registration was carried out on the AO image using the \textsc{IRAF} tasks \textsc{ccmap} and \textsc{ccsetwcs}.  The astrometric uncertainty was $\sigma_{\alpha} = 0.023\arcsec$, $\sigma_{\delta} = 0.036\arcsec$.  The position of the progenitor star in the \hst\ reference image is $\alpha= 1^{\text{h}}34^{\text{m}}14^{\text{s}}.418$, $\delta=-29^{\circ}26\arcmin23\arcsec.83$ and is detected with S/N = 5.8 for an astrometric precision of $0.052\arcsec$.  Relative astrometry from the AO image suggests that the position of SN~2016gkg is $\alpha= 1^{\text{h}}34^{\text{m}}14^{\text{s}}.424$, $\delta=-29^{\circ}26\arcmin23\arcsec.82$ for an offset of $\Delta \alpha = +0.05\arcsec$, $\Delta\delta = -0.01\arcsec$. The combined offset is well within the uncertainty from \hst\ astrometry and relative astrometry (0.89-$\sigma$; astrometric uncertainty from the position of SN~2016gkg is negligible), and we conclude that the positions of these objects agree with each other.  This evidence strongly suggests that the point source detected in archival \hst\ imaging is the progenitor star of SN~2016gkg.

We estimate the probability of a chance coincidence in the \hst\ image by noting that there are a total of $12$ point sources with S/N$>$3 in the \hst\ image from \autoref{fig:astrometry}.  The 3-$\sigma$ error ellipse for the \hst\ reference image has a solid angle of approximately 0.64$~\text{arcsec}^{2}$, which implies that $\sim7.6~\text{arcsec}^{2}$ or 0.12\% of the \hst\ archival image has a point source that is close enough to be associated with that region. This value represents the probability that the detected point source is a chance coincidence, and we find that it is extremely unlikely that the blue point source was aligned with the position of the SN by chance.

\subsection{Photometric Classification of the Progenitor Star}

From our photometric analysis of the SN~2016gkg progenitor star, we obtained ST magnitudes $m_{\text{F450W}} = 22.93\pm0.47$, $m_{\text{F606W}} = 23.40\pm0.33$, $m_{\text{F814W}} = 24.56\pm0.59$~mag.  We corrected these values for interstellar extinction using Equations (3a), (3b), (4a), and (4b) in \citet{cardelli+89} with $R_{V}=3.1$, which we use for all photometry herein.

\begin{figure}
	\includegraphics[width=0.49\textwidth]{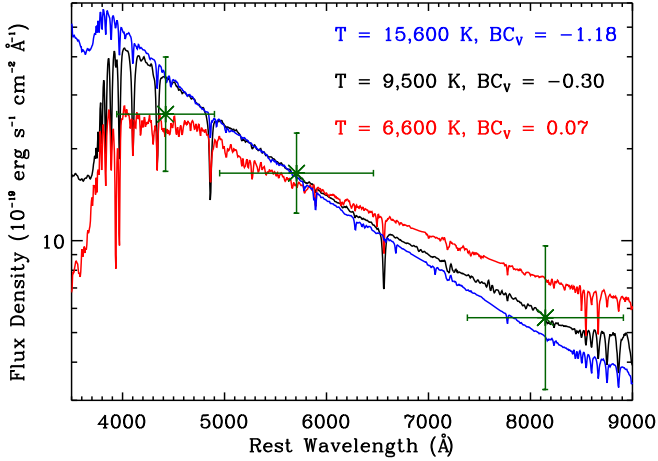}
	\caption{\hst\ archival photometry from the SN~2016gkg progenitor star.  The wavelength uncertainty of each point represents the width of the corresponding WFPC2 filter. Each point has been corrected for extinction and the recessional velocity of NGC~613 has been removed from the effective wavelength. Overplotted are the best-fit (black) and $\Delta \chi^{2}/\chi^{2}_{\text{min}} = 1$ (red/blue) stellar spectra obtained from \citet{pickles+98}. We indicate the temperature and bolometric correction of each stellar spectrum in the upper-right of the panel.}\label{fig:stellar-spectra}
\end{figure}

We used these magnitudes to determine the spectral type of the SN~2016gkg progenitor star using stellar spectra from \citet{pickles+98}. Fitting the redshift-corrected flux density to stellar spectra convolved with the WFPC2 transmission curves for the F450W, F606W, and F814W filters, we determined the best-fitting stellar spectrum and thus the temperature and bolometric correction by minimising the $\chi^{2}$ of the observed and model flux densities.  In \autoref{fig:stellar-spectra}, we show the best-fitting stellar spectrum in black along with the stellar spectra with the lowest (red) and highest (blue) implied temperatures that were within $\Delta \chi^{2} / \chi^{2}_{\text{min}} = 1$ of the minimum $\chi^{2}$.  The implied best-fitting temperature and bolometric correction are $T = 9500\substack{+6100\\-2900}~\text{K}$ and $\text{BC}_{V} = -0.30\substack{+0.37\\-0.88}$~mag. These values correspond to a spectral class of A0.

From the best-fitting stellar spectra, the implied flux density in Johnson $V$ band is $7.2\substack{+3.2\\-2.3}\times10^{-19}~\text{erg s}^{-1}~\text{cm}^{-2}$~\AA$^{-1}$ or $m_{V} = 24.3\pm0.4$~mag (implying the best-fitting stellar type is A0Ia), which suggests that the overall bolometric magnitude is $m_{\text{bol}} = 24.0\substack{+0.54\\-0.97}$~mag.  The luminosity of the SN~2016gkg progenitor star is therefore $\log (L/L_{\odot}) = 5.14\substack{+0.22\\-0.39}$ with an implied radius of $\log (R/R_{\odot}) = 2.14\substack{+0.29\\-0.59}$.  We note that these values are remarkably similar to the progenitor star model for SN~2008ax in \citet{crockett+08} where the authors found the photometry was well-fit by a B8 to early K supergiant combined with a M4 supergiant, the former having $\log (L/L_{\odot}) = 5.1, T = 8900~\text{K}$ \citep[see also][]{smartt+15}.  In our discussion of the SN~2016gkg spectrum above, we emphasize this comparison with SN~2008ax at $15~\text{days}$ after explosion.

\subsection{Matching the SN~2016gkg to Stellar Evolution Tracks}

In order to constrain the zero-age main-sequence mass ($M_{\rm ZAMS}$) and evolutionary path of a SN progenitor star, it is necessary to compare the luminosity and temperature derived from photometry to model evolutionary tracks. This analysis has been done for a number of SNe~IIb including SN~1993J \citep{podsialowski+93}, SN~2008ax \citep{crockett+08}, SN~2011dh \citep{maund+11,vandyk+11,bersten+12}, and SN~2013df \citep{vandyk+14}.  Here, we analyse the temperature and luminosity derived for SN~2016gkg to single- and binary-star models on the Hertzsprung-Russell (HR) diagram and make comparisons to these example SN~IIb progenitor stars.

\subsubsection{Single-Star Models}

\begin{figure*}
	\includegraphics[width=0.49\textwidth]{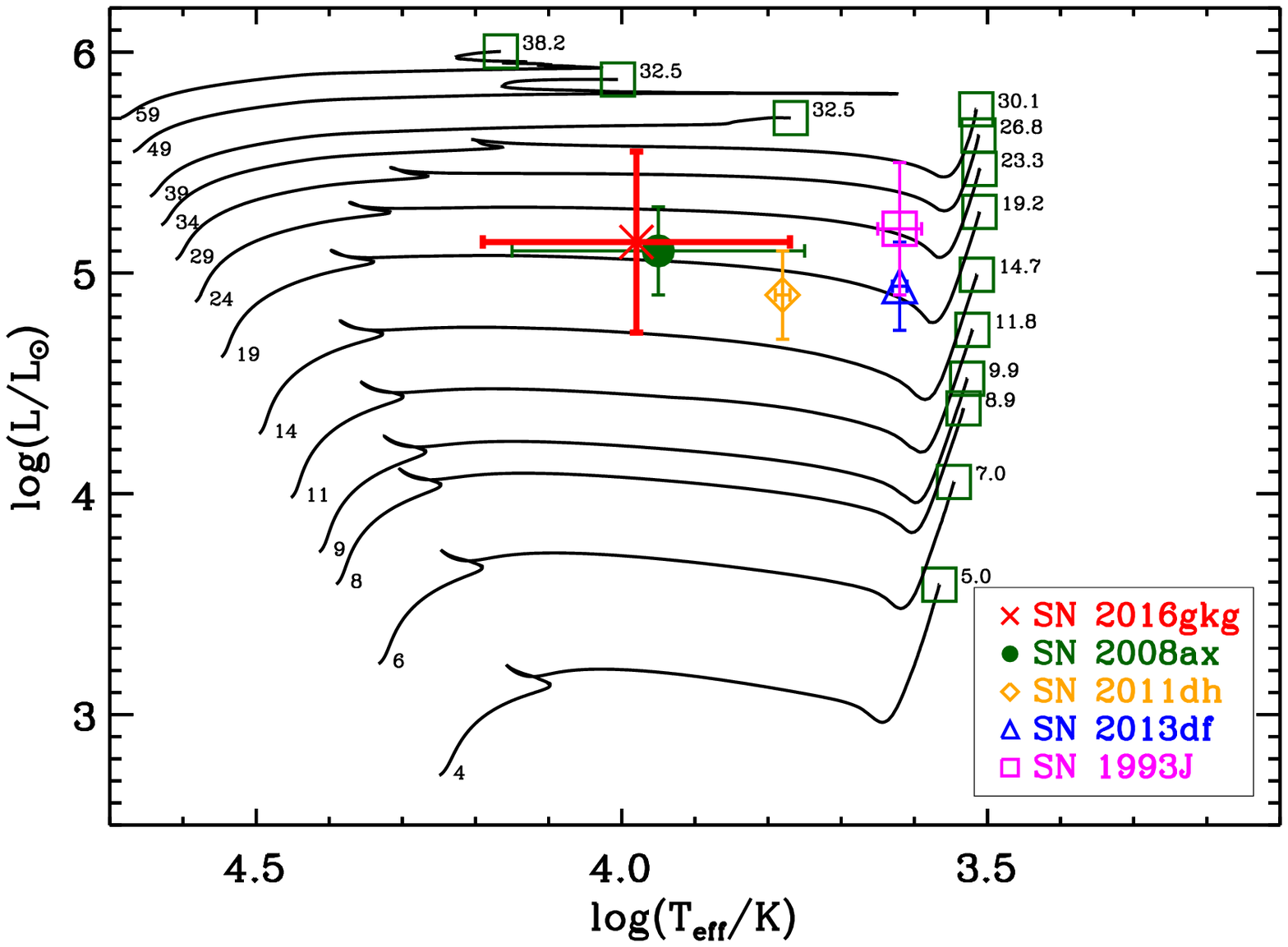}
	\includegraphics[width=0.49\textwidth]{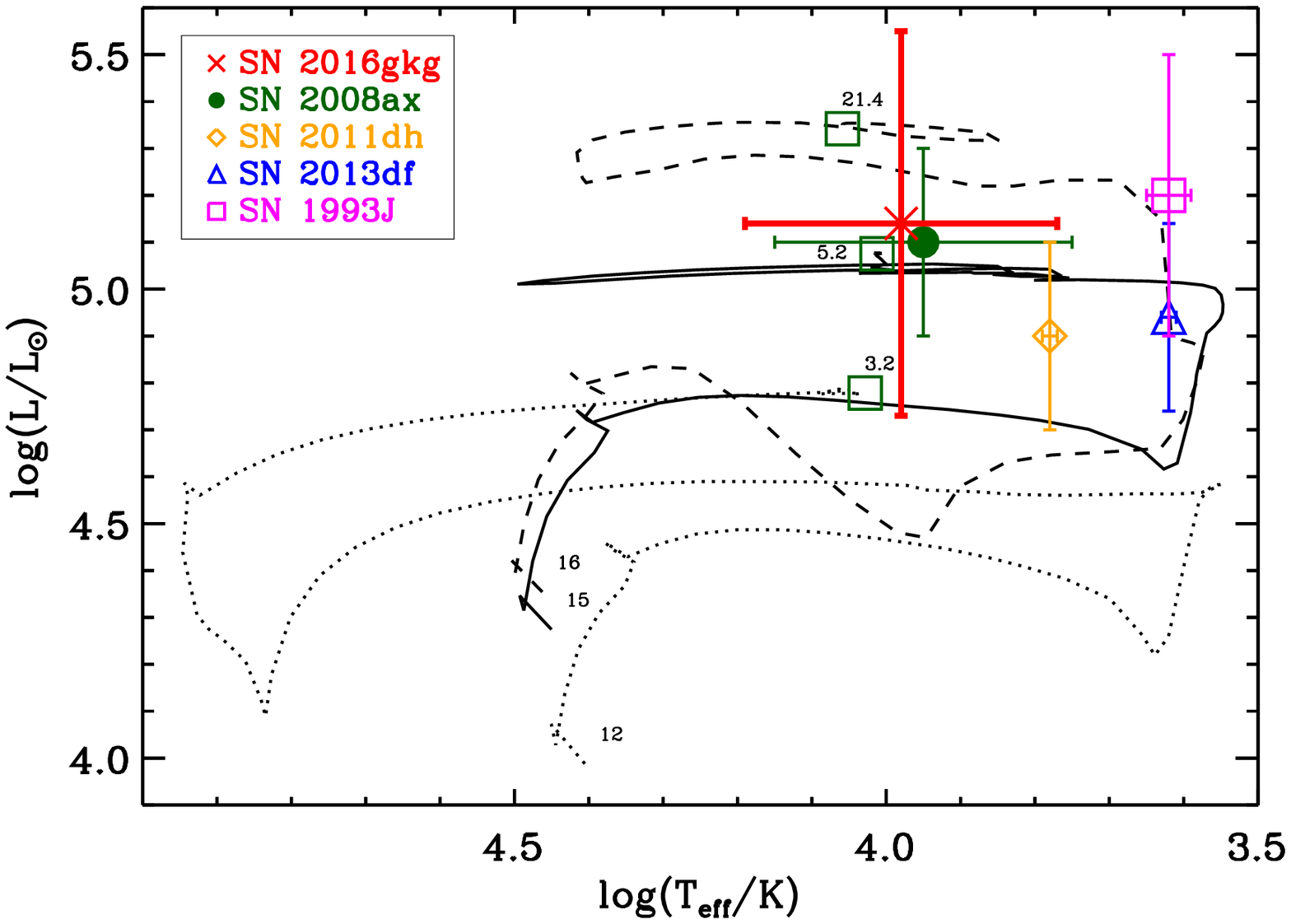}
	\caption{(Left) Single-star evolutionary tracks plotted on the HR diagram with the inferred luminosity and temperature from SN~2016gkg overplotted.  We also indicate the inferred luminosities and temperatures from the SNe~IIb 2008ax, 2011dh, 2013df, and 1993J \citep{crockett+08,maund+11,vandyk+11,vandyk+14,aldering+94}.  The initial mass and final mass of the modeled star are given near the start and end points of each evolutionary track (the latter is indicated with a square).  As we demonstrate, no single-star evolutionary track terminates near the inferred luminosity and temperature of the SN~2016gkg --- or any other SN~IIb --- progenitor star.  (Right) Same as the left but for the binary star models that terminate at values in agreement with the inferred luminosity and temperature for SN~2016gkg (as discussed in \autoref{sec:binary}).  The best-fitting model has an initial stellar mass of $M = 15~M_{\odot}$ with a $1.5~M_{\odot}$ companion.  The initial period is $1000~\text{days}$ and the primary star explodes with $M = 5.2~M_{\odot}$.  Two additional examples with initial masses $12~M_{\odot}$ ($8.4~M_{\odot}$ initial mass companion, 160 day period) and $16~M_{\odot}$ ($14.4~M_{\odot}$ initial mass companion, 6.3 day period) are shown with dotted and dashed lines, respectively. All of these models agree with the inferred luminosity and temperature of SN~2016gkg.}\label{fig:tracks}
\end{figure*}

Single-star models were obtained from \citet{brott+11} for $M_{ZAMS} = 5-60~M_{\odot}$ stars.  We examined models with metallicity $Z = 0.0088$, which was the closest set to the observed metallicity of NGC~613.  We overplot these models with the observed parameters of the SN~2016gkg progenitor star on the HR diagram in \autoref{fig:tracks}.  As we demonstrate, there are no single-star models that are consistent with ending their evolutionary tracks near the predicted luminosity and temperature values.

We find that it is extremely unlikely that SN~2016gkg originated from a single-star, even accepting moderately inflated uncertainties such that the SN~2016gkg progenitor star is consistent with stars with $M_{\rm ZAMS} \sim 40$--$50~M_{\odot}$. \citet{woosley+07} and \citet{sukhbold+16} have found that, for stars with $M_{\rm ZAMS} > 30~M_{\odot}$, the pre-supernova iron core is too large for a SN to be successful.  Moreover, mass loss is sufficiently strong that most of these stars lose their entire hydrogen envelopes and are thought to end their evolution as Wolf Rayet stars, implying that the subsequent SN would be Type Ib or Ic.  SNe~IIb require progenitor stars with extended low-mass hydrogen envelopes \citep{podsialowski+93,woosley+94,elmhamdi+06}, and any single-star model for such a system would require finely tuned mass loss that would otherwise fail to reproduce the observed range in SN~IIb light curves, spectra, and progenitor stars.  While the single-star scenario could describe a minority of SNe~IIb, it is likely that the majority of these systems come from binary-star systems such as the one observed toward SN~1993J \citep{maund+04,fox+14}.

\subsubsection{Binary-Star Models}\label{sec:binary}

We examine evolutionary tracks involving binary stars in order to assess the plausibility of these systems as possible progenitor stars for SN~2016gkg.  We obtained our binary star evolutionary tracks from the Binary Population and Spectral Synthesis (\textsc{BPASS}) code as described in \citep{eldridge+09}.  These models provide a range of metallicities ($Z=0.001$--0.040), primary star masses ($M/M_{\odot} = 0.1$--300), mass ratios ($q = 0.1$--0.9), and initial periods ($\log (P/1~\text{day}) = 0$--4).  We fixed the metallicity of the binary-star models to $Z=0.010$ in order to provide the best match to the observed metallicity of NGC~613. Otherwise, we examined the full range of parameters provided by \textsc{BPASS}.

For our fitting scheme, we looked for binary-star models that produced a primary star with terminal luminosity and temperature that matched those observed for the SN~2016gkg progenitor star.  Overall, we found $107$ out of $5565$ \textsc{BPASS} models with primary star parameters in this allowed range. In \autoref{fig:tracks}, we show the stellar evolution of the best-fitting binary-star model on the HR diagram along with the inferred luminosity and temperature of the SN~2016gkg progenitor star. The primary star has an initial mass of $15~M_{\odot}$ while the secondary (accreting) star has an initial mass of $1.5~M_{\odot}$ and an initial orbital period of $1000~\text{days}$.  We note in \autoref{fig:tracks} that the pre-explosion mass of the best-fitting star is $M = 5.2~M_{\odot}$.  The hydrogen that remains in the envelope from this best-fitting model is $5\times10^{-3}~M_{\odot}$, which agrees with models of SNe~IIb \citep{dessart+11}.

If SN~2016gkg evolved from a binary-star system, it may be possible to detect the companion star in follow-up photometry after the SN has faded. The secondary star in our best-fitting binary star model is intrinsically much fainter than the SN~2016gkg progenitor star. Accounting for distance modulus and extinction, its expected brightness in F300W is 25.9~mag. It may be feasible to search for such a companion with sufficiently deep imaging. 

\subsection{Modeling the Early-Time Light Curve of SN~2016gkg}\label{sec:light-curve}

The early-time light curve of any SN can yield important information about the progenitor star when shock break-out is observed.  For SNe other than SNe~II-P, observations of this phase are extremely scarce as their progenitor stars are thought to have less extended envelopes which implies a fast rise and decline in the early-time light curve.  In the rare cases where this phase is observed, hydrodynamical models can constrain the radius of the progenitor star, as larger stars tend to have hotter effective temperatures with a more luminous initial peak while smaller stars tend to appear cooler.  In our analysis of the early-time light curve, we use models derived from \citet[][]{rabinak+11} for a star with a hydrogen envelope density profile $\rho \approx (1 - r/R_{*})^{3}$ (where $R_{*}$ is the stellar radius). In general, we assume that the progenitor star has a blackbody color temperature 20\% larger than the photospheric temperature and typical Thomson scattering opacity $\kappa = 0.34~\text{cm}^{2}~\text{g}^{-1}$.  \citet{rabinak+11} and \citet{bersten+12} demonstrated that these assumptions are good approximations of more detailed models for $t < 1~\text{day}$ after explosion.

In order to calculate the radius of the progenitor star, we must make assumptions about the explosion energy and ejecta mass of SN~2016gkg.  These parameters are well-known for the SN~IIb 1993J, which we have demonstrated is a good match for SN~2016gkg at the epoch of our spectroscopic observation.  We employ parameters for a SN~1993J-like explosion with ejecta mass, $M_{\rm ej} = 2.6~M_{\odot}$, and explosion energy, $E = 10^{51}~\text{erg}$ \citep{woosley+94,young+95}.  Using these parameters, we fit specific luminosity to the model at a time $t$ since explosion with

\begin{eqnarray}
	L_{\lambda} &=& 0.234 \mu r^{2} \frac{\left(h c/\lambda\right)^{5}}{\exp \left(h c/\lambda T\right)-1}\\
	r &=& 3.3 \times 10^{14} \frac{E_{51}^{0.39} \kappa_{0.34}^{0.11}}{(M_{ej}/M_{\odot})^(0.28)} t_{5}^{0.78}~\text{cm} \\
	T &=& 1.6 \frac{E_{51}^{0.016} R_{*,13}^{1/4}}{(M_{ej}/M_{\odot})^{0.033} \kappa_{0.34}^{0.27}} t_{5}^{-0.47}~\text{eV}
\end{eqnarray}

\noindent where $E = E_{51} 10^{51}~\text{erg}$, $\kappa = \kappa_{0.34} 0.34~\text{cm}^{2}~\text{g}^{-1}$, $t = t_{5} 10^{5}~\text{s}$, $R_{*} = R_{*,13} 10^{13}~\text{cm}$, and $\mu = 1.14 \times 10^{12}~\text{cm}^{-3}~\text{s}^{-1}~\text{K}^{-4}$ (i.e., the ratio of the radiation constant to the Planck constant).  As we have noted, this model breaks down for times significantly (e.g., $>1~\text{day}$) after explosion.  Therefore, in determining the explosion date and stellar radius, we fit only photometry within $1.5~\text{days}$ of the ASAS-SN $V$ band limit on 20.165 September 2016.  These include the discovery magnitudes and followup photometry from \citet{atel9521} and \citet{atel9526}.

\begin{figure*}
	\includegraphics[width=0.49\textwidth]{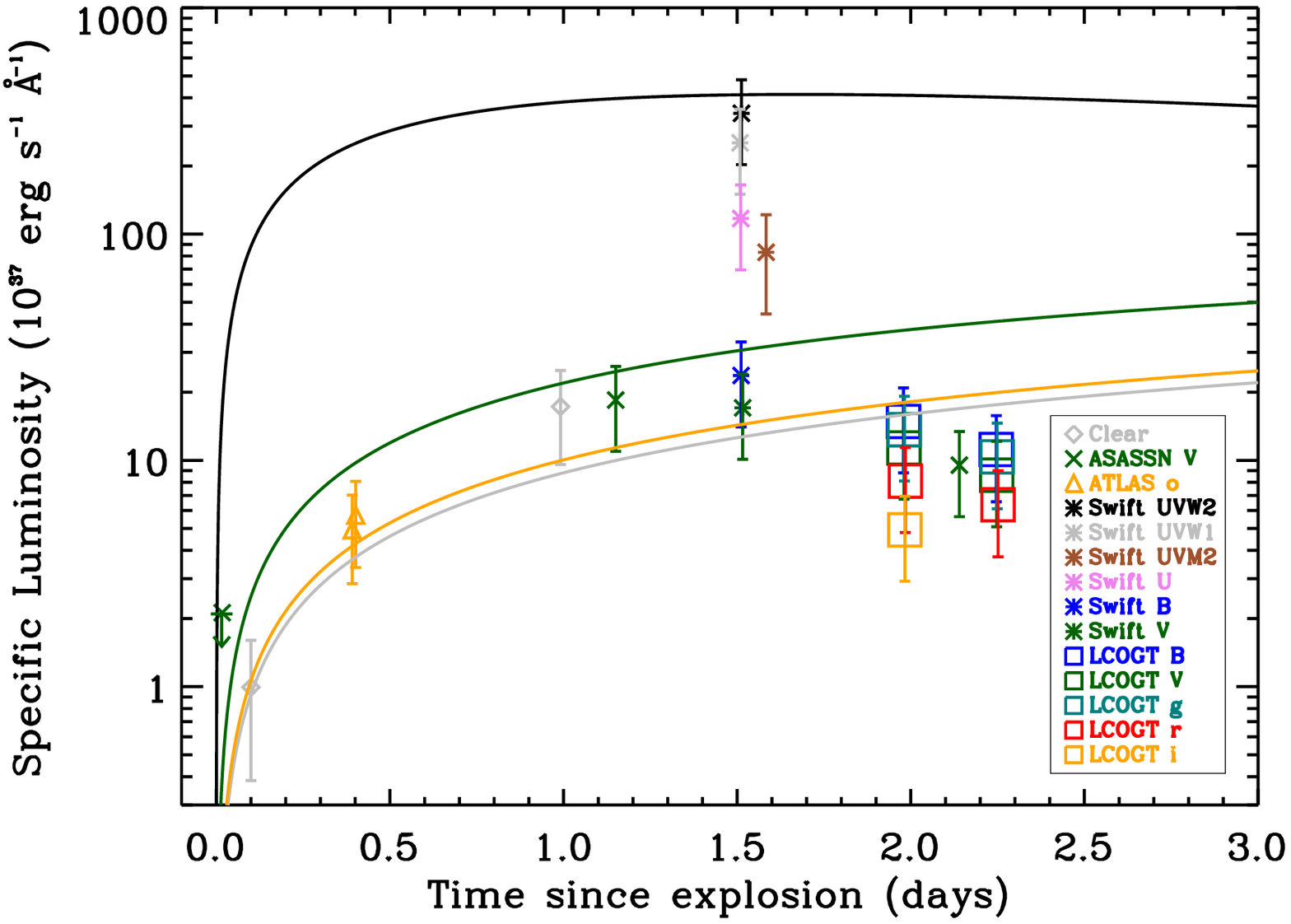}
	\includegraphics[width=0.49\textwidth]{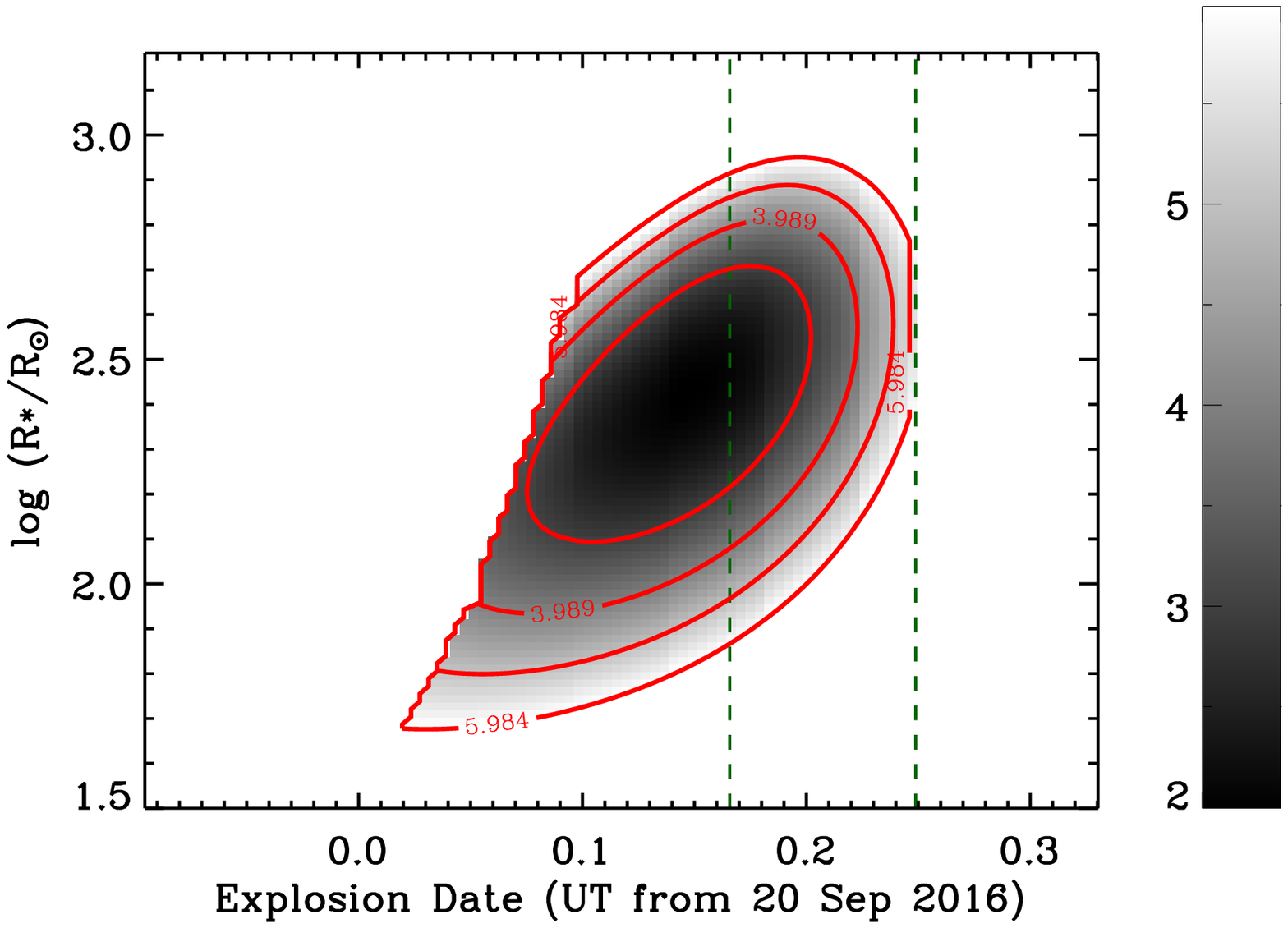}
	\caption{(Left) Early-time light curve of SN~2016gkg as discussed in \autoref{sec:light-curve} and referenced in \citet{atel9521} and \citet{atel9526}. The data are plotted in terms of specific luminosity (i.e., $L_{\lambda} = 4 \pi D^{2} f_{\lambda}$).  Overplotted are light curves of the cooling phase that follows shock break-out based on models provided in \citet{rabinak+11} and for a range of filter transmission curves including ``clear'' (grey), ATLAS $o$ (orange), ASAS-SN $V$ (green), and \swift\ UVW2 (black).  These models use the best-fitting explosion time and progenitor star radius derived from photometry within $1.5~\text{days}$ of the initial ASAS-SN $V$ band upper limit (indicated on the left with an arrow).  Other parameters used to derive these light curves are described in \autoref{sec:light-curve}. (Right) $\chi^{2}$ for the range of model parameters used to derive the light curves on the left.  We have overplotted two dashed lines to indicate the time of the ASAS-SN $V$ band upper-limit (20.1653 September 2016) and the first photometry point (20.2484 September 2016), which place the strongest constraints on the explosion date.}\label{fig:lc}
\end{figure*}

We constructed a range of models using the equations above and convolved the specific luminosity with the filter transmission curves.  In \autoref{fig:lc}, we show our best-fitting model for a range of filters, including ASAS-SN $V$, ATLAS $o$, the ``clear'' filter, and the \swift\ UVW2.  Our best-fitting model corresponds to a stellar radius of $\log (R/R_{\odot}) = 2.41\substack{+0.40\\-0.58}$ and an explosion date of $t_{0} = 20.15\substack{+0.08\\-0.10}$ September 2016.  Our range of best-fitting parameters is also displayed in \autoref{fig:lc} with contours representing $\chi^{2}$ overplotted.

The early-time light curve agrees with all of the photometry within $1.5~\text{days}$ of explosion to within the 1-$\sigma$ uncertainties.  After this point, there is general disagreement between the model and observed magnitudes, especially at redder wavelengths where the model overpredicts the specific luminosities and does not turn over as quickly as the observed light curve.  This disagreement is likely caused by our assumption of a constant Thomson-like opacity, independent of time and spatial coordinate in the model star.  In more realistic models, the opacity is sensitive to the ionisation state of the model star and decreases as hydrogen in the envelope recombines.  However, good agreement can be found at early times between this model and the SN~1987A light curve \citep{rabinak+11} and the SN~2011dh light curve \citep{bersten+12} where most of the hydrogen envelope is ionised.  Therefore, we are confident that the explosion date and progenitor star radius inferred from this model is an accurate representation of the light curve.

\section{CONCLUSIONS}\label{sec:conclusions}

We describe new astrometric and photometric analysis of the SN~2016gkg progenitor star as well as optical photometry and spectroscopy of the SN itself.  Our analysis yields new insight into SN~2016gkg and we find the following:

\begin{enumerate}

\item Astrometric analysis of our AO imaging indicates the SN position is consistent with the position of a blue point source in \hst\ imaging.  Fitting the photometry of this source, we find the best-fitting stellar model to be an A0Ia star with $\log (L/L_{\odot}) = 5.14\substack{+0.22\\-0.39}$ and $T = 9500\substack{+6100\\-2900}~\text{K}$ and implied radius of $\log (R/R_{\odot}) = 2.14\substack{+0.29\\-0.59}$.

\item Based on the best-fitting luminosity and temperature of the SN~2016gkg progenitor star, we find that single-star models do not terminate with the inferred properties.  Rather, we find that binary-star models are required to produce evolutionary tracks with primary star terminal properties that match the SN~2016gkg progenitor star. The best-fitting binary-star model involves a primary star with $M_{\rm ZAMS} = 15~M_{\odot}$ and a secondary with $M_{\rm ZAMS} = 1.5~M_{\odot}$. With sufficiently deep imaging, it may be possible to detect the secondary star once the SN has faded significantly.

\item We fit analytic models of the cooling phase that follows shock break-out to the specific luminosity observed from SN~2016gkg.  These models are sensitive to both the explosion date and radius of the progenitor star.  Our best-fitting explosion date and progenitor star radius are $t_{0} = 20.15\substack{+0.08\\-0.10}$ September 2016 and $\log (R/R_{\odot}) = 2.41\substack{+0.40\\-0.58}$.  The latter value is in agreement with the radius fit to the progenitor star from pre-explosion photometry.

\end{enumerate}

\smallskip\smallskip\smallskip\smallskip
\noindent {\bf ACKNOWLEDGMENTS}
\smallskip
\footnotesize

We would like to thank John Tonry for providing data from ATLAS, as well as Dan Kasen, Josiah Schwab, Enrico Ramirez-Ruiz, and Stan Woosley for their helpful discussions. We also thank Marc Kassis for his assistance with Keck data acquisition and Nathan Smith and Jennifer Andrews for discussing spectroscopy of SN~2016gkg.

The UCSC group is supported in part by NSF grant AST-1518052 and from fellowships from the Alfred P.\ Sloan Foundation and the David and Lucile Packard Foundation to R.J.F.

Some of the data presented herein were obtained at the W. M. Keck Observatory, which is operated as a scientific partnership among the California Institute of Technology, the University of California, and NASA. The observatory was made possible by the generous financial support of the W. M. Keck Foundation. We wish to recognise and acknowledge the cultural significance that the summit of Mauna Kea has within the indigenous Hawaiian community. We are most fortunate to have the opportunity to conduct observations from this mountain.

Some of our photometry of SN~2016gkg comes from All Sky Automated Survey for SuperNovae (ASAS-SN), the Asteroid Terrestrial-impact Last Alert System (ATLAS), \swift, and the Las Cumbres Observatory Global Telescope Network (LCOGT). The \textit{Hubble Space Telescope} (\hst) is operated by NASA/ESA. Some of our analysis is based on data obtained from the \hst\ archive operated by STScI. Our analysis is based in part on observations obtained at the Southern Astrophysical Research (SOAR) telescope, which is a joint project of the Minist\'{e}rio da Ci\^{e}ncia, Tecnologia, e Inova\c{c}\~{a}o (MCTI) da Rep\'{u}blica Federativa do Brasil, the U.S. National Optical Astronomy Observatory (NOAO), the University of North Carolina at Chapel Hill (UNC), and Michigan State University (MSU).

\textit{Facilities}: Keck (NIRC2), SOAR (Goodman)

\bibliography{2016gkg}

\end{document}